\documentstyle[11pt,epsf]{article}

\newcommand{\be}{\begin{equation}}   \newcommand{\ee}{\end{equation}}
\newcommand{\bear}{\begin{eqnarray}}
\newcommand{\eear}{\end{eqnarray}}
\newcommand{\ba}{\begin{array}}      \newcommand{\ea}{\end{array}}

\newcommand{\drawsquare}[2]{\hbox{%
\rule{#2pt}{#1pt}\hskip-#2pt
\rule{#1pt}{#2pt}\hskip-#1pt
\rule[#1pt]{#1pt}{#2pt}}\rule[#1pt]{#2pt}{#2pt}\hskip-#2pt
\rule{#2pt}{#1pt}}
\newcommand{\Yfund}{\raisebox{-.5pt}{\drawsquare{6.5}{0.4}}}
\newcommand{\Ysymm}{\raisebox{-.5pt}{\drawsquare{6.5}{0.4}}\hskip-0.4pt%
        \raisebox{-.5pt}{\drawsquare{6.5}{0.4}}}
\newcommand{\Yasymm}{\raisebox{-3.5pt}{\drawsquare{6.5}{0.4}}\hskip-6.9pt%
        \raisebox{3pt}{\drawsquare{6.5}{0.4}}}

\def\vbr{$\vphantom{\sqrt{F_e^i}}$}
\begin{document}

\twocolumn[ 
{\small BUHEP-97-12 \hspace{12.9cm}  hep-ph/9703390}\\
\begin{center}
\vspace{-2mm}
{\LARGE {\bf $B - L$ Mediated Supersymmetry Breaking}} \\
\vspace{7mm}
{\Large Bogdan A. Dobrescu} \footnotemark\\
\vspace*{0.2cm}
{\it Department of Physics, Boston University, Boston, MA 02215, USA}\\
\vspace*{0.4cm}
March 21, 1997 \ (revised April 29, 1997)
\end{center}
\begin{quote}
I present a realistic model of dynamical supersymmetry breaking,
in which a $U(1)_{B-L}$ gauge interaction communicates supersymmetry
breaking to the standard fields.
A distinctive superpartner spectrum is predicted in this model.
\end{quote}
\vspace{3mm}
]
\baselineskip=18pt
\pagestyle{plain}
\footnotetext{e-mail address: dobrescu@budoe.bu.edu}

\subsection*{1. \ Motivation}
\vspace{-2mm}
The large hierarchy between the electroweak scale, $v$, 
and the Planck scale
indicates the existence of a symmetry which protects $v$ against
quadratic divergences, and is dynamically broken.
If there are light fundamental scalars, such as the Higgs doublet,
then the ``protective'' symmetry must be supersymmetry.

If the superpartners of the standard model fields find out about 
supersymmetry breaking from the standard gauge interactions, then the
superpartner spectrum can be computed in terms of few parameters,
and unwanted flavor-changing neutral currents are suppressed.
This standard gauge mediated supersymmetry breaking (SGMSB)
scenario requires ``messenger'' superfields charged under 
the standard gauge group,  with a nonsupersymmetric spectrum.
In the usual SGMSB models \cite{dns}, where the 
vacuum expectation values (VEV's) of the scalar-
and $F$-components of a gauge singlet superfield give 
the masses of the messenger fields, only false vacua have 
experimentally viable properties \cite{noi,berk}.
The true vacuum may become viable 
if a second gauge singlet is included \cite{noi,berk}, 
but in this case there are many independent parameters.
If mass terms for the messenger fields are included in the
superpotential, then simple SGMSB models
can be constructed \cite{mass}.
It remains, though, to be shown that this simplicity is preserved
once a dynamical origin for these masses is specified.
Other recent SGMSB models involve non-generic and nonrenormalizable 
interactions \cite{nonren}.

A different possibility is that the 
messenger of supersymmetry breaking is a 
spontaneously broken gauge interaction.
The simplest choice is a $U(1)_{B-L}$ that couples to the $B - L$
number. 
The possibility of using $U(1)_{B-L}$
as messenger was first suggested in ref.~\cite{ads1}, but
model building efforts in this direction have been hampered by several
phenomenological problems: 
{\it i)} 
it is difficult to give rise to positive squared-masses for squarks
and sleptons;
{\it ii)} the usual gauginos do not couple to $B - L$ so that they
remain too light;
{\it iii)} 
a natural mechanism of breaking $U(1)_{B-L}$ spontaneously should
be found.
A model which uses a combination of $U(1)_{B-L}$ and 
hypercharge as messenger is presented in ref.~\cite{bml}. However, 
this is not a model of dynamical supersymmetry breaking (DSB)
because supersymmetry breaking is introduced through Fayet-Iliopoulos
terms.

Here I construct a renormalizable
DSB model with $U(1)_{B-L}$ as messenger,
which solves the phenomenological problems listed above, and as a
consequence predicts a peculiar superpartner spectrum.

\subsection*{2. \ The Model}
\label{sec:color}
\setcounter{equation}{0}
\renewcommand{\theequation}{2.\arabic{equation}}
\vspace{-2mm}
In addition to the minimal supersymmetric standard model (MSSM) fields,
the model I propose contains an $SU(5) \times SU(2) \times U(1)_{B-L}$
gauge group and the chiral superfields shown in table 1, which are
singlets under the standard model gauge group.

\begin{table}[htbp]
\centering
\begin{tabular}{|c|c|c|c|}\hline
 & $SU(5)$ & $SU(2)$ & $U(1)_{B-L}$ \vbr
\\\hline \hline
$A_1, A_2$ & \Yasymm & {\bf 1} & 0 \vbr\\  \hline
$B$ &  $\overline{\Yfund}$ & \Yfund & 0 \vbr \\ \hline
$\phi$ & {\bf 1} & \Ysymm & 0 \vbr\\ \hline
$\chi_{0,\pm}$ & {\bf 1} & \Yfund & $0, \, \pm y$ \vbr\\ \hline
\end{tabular}
\label{Sp}
\parbox{3in}{\caption{Field content of the model.}}
\end{table}

The most general dimension-3 terms in the superpotential are given by
\be
W = \lambda_0 A_1 B^2 + \lambda \chi_0^2 \phi + 
\kappa\chi_+\chi_-\phi ~.
\ee 
To avoid a hierarchy problem, the mass terms $\phi^2$ and
$\chi_+\chi_-$ are excluded by invoking a discrete symmetry.

The $SU(2)$ group is infrared free, so that it must be 
in the weak coupling regime.
The $SU(5)$ instantons generate the following effective superpotential
\cite{su5}:
\be
W_{\rm np} = \frac{\Lambda_5^{11}}{(A_1 A_2)^3 B^2} ~,
\ee
where $\Lambda_5$ is the scale of $SU(5)$.

The scalar potential can be written as
\be
V = V_5 + V_2 ~,
\ee
\vspace*{-7mm}\\
where 
\be
V_5 = \frac{g_0^2}{2} D_5^{\hat{a}} D_5^{\hat{a}} 
+ |F_A|^2 + |F_B|^2
\ee
is the potential of the $SU(5)$ DSB model with two generations
\cite{su5},
with $D_5^{\hat{a}}$ the $D$-term of $SU(5)$, and $F_A, F_B$ the
usual $F$-terms of the $A$ and $B$ fields,
while
\be
V_2 = \frac{g^2}{2} D_2^a D_2^a + |F_\phi|^2 + |F_\chi|^2 
\ee
includes the $SU(2)$ $D$-potential and the remaining $|F|^2$ terms.

The most general parametrization of the $B$
scalar field, up to an $SU(5) \times SU(2)$ transformation,
is given by
\be
B = \left(\ba{ccccc} b & 0 & 0 & 0 & 0\\ [1mm]
                     0 & b^\prime & 0 & 0 & 0\\ 
        \ea\right)~.
\label{unitar}
\ee
At the minimum at least one of $b$ and $b^\prime$ is nonzero
(otherwise $F_A \rightarrow \infty$), so that $SU(2)$ is completely
broken. 

At scales where the $SU(5)$ gauge coupling $g_0$ is much larger than 
$\lambda_0$, the global minimum of $V_5$ lies along the $D_5$ flat
directions. In this case, i.e. to leading order in $\lambda_0^2/g_0^2$,
it has been shown numerically \cite{veld} that
a flavor-$SU(2)$ symmetry of the $SU(5)$ DSB model
is preserved, which requires
$|b| = |b^\prime| > 0$ at the minimum. Thus, the VEV of $B$ does not
contribute to the $SU(2)$ $D$-term to leading order in
$\lambda_0^2/g_0^2$.
As a result, the deepest minimum of $V$ corresponds
to the deepest minima of both $V_5$ and $V_2$.
Because $V_2$ has many classically flat directions, the vacuum energy
is given entirely by $V_5$.
At the minimum $b \sim \lambda_0^{-1/11} \Lambda_5$
and the vacuum energy is of order $\lambda_0^{1/2} b$, so that
supersymmetry is spontaneously broken.
Therefore,
the flat directions of $V_2$ are lifted by radiative corrections.

The $\phi$ and $\chi$ scalars get masses at one-loop.
In the Landau gauge, the only relevant diagrams are the ones shown in
fig.1. 
The computation of these diagrams is difficult because there is mixing 
between the $A_1$, $A_2$ and $B$ states. It is possible, however, to
estimate 
the one-loop masses of the $\phi$ and $\chi$ scalars by adapting to
the present $SU(2)$ sector the 
Feynman rules given in ref.~\cite{guha} for the Higgs sector of the
MSSM.
This amounts to compute only the contributions from the $SU(2)$
doublets which acquire VEV's, i.e. $B_1$ and $B_2$, where 1 and 2 are
$SU(5)$ indices [see eq.~(\ref{unitar})].
\begin{figure}[! t]
\vspace*{-3.1cm}
{\hbox{\epsfxsize=10cm\epsfysize=8cm \epsfbox{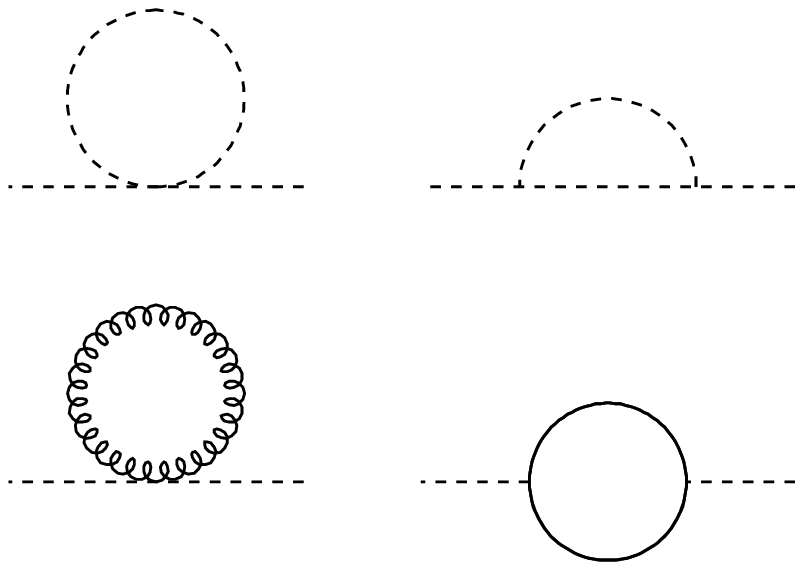}}}

\vspace{-.3mm}\noindent
\makebox[1cm][l]{{\small Fig.~1.}}
\parbox[t]{7.55cm}{ 
{\small Masses for the
$\phi$ and $\chi$ scalars (horizontal lines) from interactions with
$B$ scalars (upper graphs), $SU(2)$ gauge bosons and $SU(2)$
gauginos.} }
\vspace{-1mm}\end{figure}
The result is finite and negative:
the one-loop squared-mass of a scalar in the $R$ representation 
of $SU(2)$ is given by 
\be
M^2_R = -  C_R  M^2 ~,
\label{oneloopmass}
\ee
where $M^2 > 0$ and $C_R$ is the eigenvalue of the Casimir 
operator (3/4 for the doublet).
$M^2$ has a simple expression if expanded
in powers of $\lambda_0^2/g^2$. The leading term, proportional to $g^4$,
cancels because supersymmetry is exact in the $\lambda_0
\rightarrow 0$ limit. The next term in the expansion is positive:
\be
M^2 = \frac{g^2}{8 \pi^2}\left(\mu_0^2 + m_0^2\right) + 
{\cal O} (\lambda_0^4) ~.
\label{msquare}
\ee
Here $\mu_0 > 0$ is the 
analogue of the $\mu$-term from the Higgs sector, i.e.
the coefficient of the $B_1 B_2$ fermion mass term ($\mu_0 \sim
\lambda_0 b$),
and $m_0^2 > 0$ is the coefficient of the $B_1 B_2$ scalar mass term
($m_0 \sim \lambda_0 b$).
The mixing between the $A_1$, $A_2$ and $B$ states will change the
numerical
coefficient in eq.~(\ref{msquare}), but it appears reasonable to assume
that the sign of $M^2$ will not change.

With the $\phi$ and $\chi$ negative mass terms included, the scalar 
potential
\be
V_2^\prime \equiv V_2 - M^2 \left[ 2 |\phi|^2 +
\frac{3}{4} \left( |\chi_0|^2 + |\chi_\pm|^2 \right)\right] 
\ee
has a runaway direction: $V_2^\prime \rightarrow - \infty$ for
$\chi_{0,\pm} = 0$ and $|\phi| \rightarrow \infty$.
However, the full scalar potential $V$ is positive definite, which
implies that the runaway direction is lifted by higher dimensional
terms. For example,
one-loop diagrams similar to the ones in fig.1 but with four external
legs induce a $|\phi|^4$ term in the effective potential, with a
coefficient of order $g^2 \lambda_0^2/(4\pi)^2$. 
There is also an
infrared divergent contribution to the $|\phi|^4$ term which should be
eliminated by summing up the complete one-loop effective potential,
leading to a $\phi^2\,$log$\phi$ term.
The balance between the $|\phi|^2$ term and
the higher dimensional terms gives the global minimum at
\be
\chi_{0,\pm} = 0 \; ,  \;  \; |\phi| \sim b ~.
\label{min}
\ee
The soft supersymmetry breaking terms generated in the MSSM
(see the following sections) are only logarithmically sensitive to the 
value of $|\phi|$ because this VEV gives supersymmetric contributions
to the $\chi_\pm$ masses.
Note that the VEV of $\phi$ breaks a global $U(1)$ and the
resulting Goldstone boson, which has anomalous couplings to the
$SU(2) \times U(1)_{B-L}$ gauge bosons, is likely to get a 
Planck scale suppressed mass, 
of order $|\phi|^{3/2} M_{P}^{-1/2}$.

\vspace*{-1mm}
\subsection*{3. \ Squark and slepton spectrum}
\label{sec:3   }
\setcounter{equation}{0}
\renewcommand{\theequation}{3.\arabic{equation}}
\vspace{-2mm}
In the vacuum (\ref{min}) all the $\chi$ and $\phi$ fields are massive.
The four scalar components of the $\chi_\pm$ superfields
are degenerate, with mass
\be
M^2_{\chi_\pm} = \frac{\kappa^2}{2} |\phi|^2 - \frac{3}{4} M^2 ~.
\ee
The fermion components of $\chi_\pm$ form
two degenerate Dirac fermions of mass
\be
m_{\chi} = \frac{\kappa}{\sqrt{2}} |\phi| ~.
\label{chif}
\ee
Because the spectrum of $\chi_\pm$ is nonsupersymmetric,
the squarks and sleptons, as well as any other  
scalar charged under $U(1)_{B-L}$, get masses
at two loops (the one-loop 
contributions from the $\chi_+$ and $\chi_-$ scalars cancel each
other). The leading logarithmic term is given by \cite{gmass}
\be
\hspace*{-3mm}
M^2_{B-L} = - \left[(B-L) y {\alpha_{\scriptscriptstyle B-L} \over
  2 \pi} \right]^2 {\rm Str} (M^2_{\chi}) 
\ln\left(\frac{\Lambda}{m_{\chi}}\right),
\label{sum}
\ee
where the cut-off is of the order of the 
$SU(5)$ gauge boson mass, $\Lambda \sim g_0 b$, and
the supertrace over the eight $\chi_\pm$ states,
\be 
{\rm Str} (M^2_{\chi}) = - 3 M^2 ~,
\ee
is negative because of the $SU(2)$-induced one-loop mass [see
eq.~(\ref{oneloopmass})] of the $\chi_\pm$ scalars.

Thus, the squark and slepton squared-masses 
are indeed positive, which is the primary motivation 
for the choice of the field content in table 1. 
Given that the squarks have baryon number $\pm 1/3$ and 
the sleptons have lepton number $\mp 1$, 
from eq.~(\ref{sum}) follows an interesting 
prediction for the relation between the slepton mass, $m_{\tilde{L}}$,
and the squark mass, $m_{\tilde{Q}}$:
\be
m_{\tilde{L}} = 3 m_{\tilde{Q}} ~.
\label{prediction}
\ee
This prediction is in contrast with the ones from 
SGMSB models and from supergravity scenarios \cite{snowmass}, 
where the squarks are heavier than the sleptons.
Hence, if the squarks and sleptons will be observed and their
masses measured, then 
the prediction (\ref{prediction}) will be an important test 
of the model presented here.
Another feature is that all the sleptons are degenerate
(the electroweak corrections are negligible),
whereas in SGMSB models the left-handed sleptons are few times 
heavier than the right-handed ones.
As discussed in the next section, additional fields are 
necessary for producing gaugino masses. These fields will 
contribute negatively to the squark squared-masses, so that the
slepton-to-squark mass ratio increases further.
Note that the squark degeneracy is slightly lifted by 
electroweak corrections from the additional fields, and by the
stop mixing due to the top Yukawa coupling.

Another difference from the SGMSB models is that
the Higgs scalars, $H_u$ and $H_d$, do not get masses at two loops
because they do not carry $B - L$ charge.
However, they get masses at three loops from the interactions with
the squarks and sleptons. These masses can be estimated by integrating
out the $\chi$ fields and computing one-loop radiative corrections
in the effective theory of heavy squarks and sleptons. Because of the
large top Yukawa coupling, the stop-loop gives a negative squared-mass
to $H_u$ which drives electroweak symmetry breaking, like in the SGMSB
case.
In order to set $v \sim$ 246 GeV without
fine-tuning,
the squark mass should be of order few hundred GeV.
The slepton mass, in turn, is of order 1 TeV.

The squark mass $m_{\tilde{Q}}$ being roughly known, the value of $b$
can be computed
from eqs.~(\ref{sum}) and (\ref{msquare}), and then 
one can find out the vacuum energy. 
Some typical values of the parameters, 
$y, g \sim 1$, $\alpha_{B-L} \sim 10^{-2}$, 
$g_0/\kappa \sim 4\pi$, yield a vacuum energy of order $10^3$ TeV,
which corresponds to a gravitino mass of few hundred eV. 
In this case, 
the lightest standard model superpartner could decay 
within the detector \cite{goldstino},
and the gravitinos may be a
dark-matter component  \cite{gravitino}.

\vspace*{-1mm}
\subsection*{4. \ Gaugino masses and $U(1)_{B-L}$ breaking}
\label{sec:4   }
\setcounter{equation}{0}
\renewcommand{\theequation}{4.\arabic{equation}}
\vspace{-2mm}
The gauginos of the standard gauge group do not couple to 
$U(1)_{B-L}$, so that their masses arise only at two loops and are 
of order 1 GeV. Although such light gauginos are not conclusively 
ruled out \cite{bml,gauginos}, 
it appears more plausible that the gaugino masses
are of the order of the electroweak scale.
Another problem of the model presented in section 2 is that
$U(1)_{B-L}$ is unbroken.
These two phenomenological problems can be solved by extending
the chiral content of the model.

Consider two chiral superfields, $q$ and $\bar{q}$, belonging to some 
vector-like representation of the standard gauge group,
and three chiral superfields, $S_{0,\pm}$, which carry $B-L$ charges 
$0,\pm y_S$, and are singlets under the standard gauge group.
The superpotential is given by
\be
W = \eta \bar{q} q S_0 + \frac{\xi}{3} S_0^3 + \zeta S_0 S_+ S_- ~.
\ee
The other dimension-3 terms, $S\phi^2$ and $S \chi_+ \chi_-$,
can be eliminated by discrete symmetries.
The simplest choice of vector-like representation that preserves the
gauge coupling unification
is $q + \bar{q} \in  n \times$({\boldmath $5 + \overline{5}$})
of the grand unified $SU(5)_{\rm SM}$ group. 
Gauge coupling perturbativity up to the unification scale 
requires $n \leq 4$.
If these superfields carry $B-L$ charges $\pm y_q$, then their scalar
components get real masses equal to $3 y_q m_{\tilde{Q}}$.
Consequently, the scalar $S_0$ receives a negative squared-mass
via the one-loop diagram shown in fig. 2,
\be
m_{S_0}^2 = - 5 n \left( \frac{3 y_q m_{\tilde{Q}} \eta
  }{2\pi}\right)^2
\ln\left( \frac{\Lambda^{\prime}}{3 y_q m_{\tilde{Q}}} \right)
~,
\ee
where $\Lambda^{\prime}$ is a cut-off of order $m_\chi$.
As a result, $S_0$ acquires a VEV of order $m_{S_0}$, while
an $F_{S_0}$-term of order $m_{S_0}^2$ is induced.
Therefore, the  $q$ and $\bar{q}$ scalars end up with diagonal 
squared-masses equal to $(3 y_q m_{\tilde{Q}})^2 + \eta^2 S_0^2$,
and off-diagonal squared-masses equal to $\eta F_{S_0}$, while the
$q$ and $\bar{q}$ fermions pair and get Dirac masses 
equal to $\eta S_0$, so that the usual gauginos
receive masses at one-loop \cite{gmass}:
\be
m_{\tilde{g}_i} = n \eta S_0 \frac{\alpha_i}{4 \pi} \left|
\frac{r_1 \ln r_1}{1-r_1} - \frac{r_2 \ln r_2}{1-r_2} \right| ~,
\label{gauginomass}
\ee
where $\alpha_i$ ($i = 1,2,3$) are the 
$SU(3)_C \times SU(2)_W  \times U(1)_Y$ coupling constants, 
with the usual $SU(5)_{\rm SM}$ normalization of the hypercharge 
coupling constant [$\alpha_1 = (5/3) \alpha_Y, \, Y = 2(Q - T_3)$],
and
\be
r_{1,2} = \left( \frac{3 y_q m_{\tilde{Q}}}{\eta S_0}\right)^{\! 2}
+ 1 \pm \frac{F_{S_0}}{\eta S_0^2} ~.
\ee
\begin{figure}
\vspace*{-1cm}
{\hbox{\epsfxsize=20cm\epsfysize=25cm \epsfbox{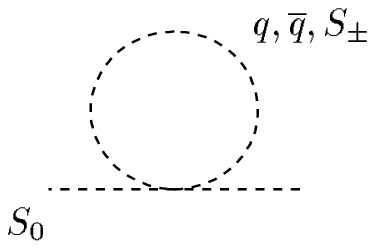}}}
\vspace*{-21.7cm}
\noindent\makebox[2.5mm][l]{ }
\parbox[t]{8.2cm}{ {\small Fig.~2. \
Negative squared-mass for  the $S_0$ scalar.
} }
\vspace{1mm}
\end{figure}
Acceptable gaugino masses can be produced for a range of parameters.
For example, the values
\be
m_{\tilde{Q}} \sim 500 \, {\rm GeV}, \;
y_q \sim 3, \; \eta \sim 1.5, \; n = 3 ~, 
\label{values}
\ee
which correspond to $|m_{S_0}| \sim 10$ TeV, yield
a gluino mass of about 200 GeV.
The ratios between the three Majorana gaugino masses 
differ from the ones given by eq.~(\ref{gauginomass})
if $q$ and $\bar{q}$ belong to other representations of the
standard model gauge group \cite{general}. 

The $q$ and $\bar{q}$ fields have also contributions at two loops
to the squark and slepton squared-masses, which are given by
eq.~(\ref{sum}) with the coupling constants, charges and masses
appropriately replaced \cite{gmass}. 
These contributions are negative because 
the $q$ and $\bar{q}$ scalars are heavier than their fermion
partners. For example, the values (\ref{values}) give a decrease in
$m_{\tilde{Q}}$ of
about 50\%, while the effect on the slepton mass is negligible.

For $R \equiv \zeta/ \xi < 1$
and $R(1-R) m_S^2 > 2 (3 y_S m_{\tilde{Q}})^2$,
the $S_\pm$ scalars acquire VEV's too, breaking $U(1)_{B-L}$ as
required, at a scale of order 10 TeV.
Note that the $D$-term for $U(1)_{B-L}$
cancels (because $S_+ = S_-$ at the minimum), so that 
there are no problems with kinetic mixing between the $U(1)_{B-L}$
gauge boson and the hypercharge gauge boson \cite{kin}.

A mechanism similar to the one
described above can be used to generate 
the $\mu$- and $B$-terms.
For this purpose there is need for three new standard model singlets
$S_{0,\pm}^\prime$, which carry $B-L$ charges 
$0,\pm y_S^\prime$. 
For $y_S^\prime < y_S$, a VEV for $S_{0}^\prime$ in the 
few hundred GeV range can be produced, which then gives the 
higgsino mass and the $B$-term via an $S_0^\prime H_u H_d$
term in the superpotential.
Other mechanisms for generating the $\mu$- and $B$-terms
may also be used \cite{mu}.

It should also be mentioned that the $U(1)_{B-L}$ anomaly cancellation
requires right-handed neutrinos. The VEV of $S_+$ can be used to
generate a Majorana mass for the right-handed neutrinos, such
that small neutrino masses arise by the see-saw mechanism.
One has to worry though that the position of the vacuum may be changed
if an $S_+ \nu^c \nu^c$ term is included in the superpotential.
Alternately, neutrino masses may be prevented by discrete
symmetries (in this case the lower bound on the $U(1)_{B-L}$ gauge 
boson's mass set by primordial nucleosynthesis is approximately 2 TeV
\cite{apj}).

\vspace*{-1mm}
\subsection*{5. \ Outlook}
\label{sec:conc}
\setcounter{equation}{0}
\renewcommand{\theequation}{5.\arabic{equation}}
\vspace{-2mm}
The model described here contains a rather large number of parameters:
6 Yukawa couplings and 3 gauge couplings in addition to the 
standard model. It is also unsatisfactory that the fields which
induce squark and slepton masses cannot be used to produce directly
the gaugino masses and to break $U(1)_{B-L}$. Furthermore, a separate
sector should be introduced for producing the $\mu$-term.
However, this model is more economical than the other known
viable DSB models. For example, the simplest complete
model of SGMSB with a viable true vacuum \cite{noi},
contains 13 parameters in the superpotential, 3 gauge couplings,
and the sector that produces the $\mu$-term, in addition to the
standard model parameters. 
Nevertheless, it would be desirable to find a common origin for the 
sectors that are responsible for gaugino and scalar masses, such as
a grand unified theory in which the fields from different sectors 
belong to the same representation.

Another unpleasant feature of the SGMSB models 
shared by the model proposed here is that some renormalizable terms
should be eliminated from the superpotential by 
discrete symmetries. It is unclear whether such symmetries
are not badly violated by Planck scale effects. And in case they are 
preserved, one has to ensure that the domain walls can 
decay, or that a period of late inflation is possible.
Inflation may be also needed for diluting 
the lightest $q-\bar{q}$ state \cite{mesfield}.

Despite these drawbacks,
the model proves the possibility of $B-L$ mediated
supersymmetry breaking.
Its importance lies in the distinctive
predictions for the superpartner spectrum, most notably being
the large ratio ($\geq 3$) between the slepton and squark masses.
Finally, it is worth pointing out that there are other light neutral
states besides the gravitino and the Goldstone boson discussed in
section 2: the DSB sector contains a massless 
fermion \cite{veld}, and an
$R$-axion \cite{raxion} with a mass of order 100 MeV given by 
supergravity effects.

\vspace*{-1mm}
\subsection*{{Acknowledgements}}
\vspace*{-3mm}
I would like to thank Sekhar Chivukula, Indranil Dasgupta, 
Benjamin Grinstein, Lisa Randall, and Martin Schmaltz for useful
discussions, and Tonnis ter Veldhuis for helpful explanations
regarding ref.~\cite{veld}.
I am grateful to Csaba Csaki and Witold Skiba for pointing out an
error in an earlier version of this paper.
This work was supported in part by the National Science
Foundation under grant PHY-9057173, and by the Department of Energy
under grant DE-FG02-91ER40676.


\vfil

\small
\begin{thebibliography}{99}

\bibitem{dns} M. Dine, A. E. Nelson and Y. Shirman, Phys.\ Rev.\ {\bf
    D 51} (1995) 1362, hep-ph/9408384; \\
  M. Dine, A. E. Nelson, Y. Nir and Y. Shirman,
    Phys.\ Rev.\ {\bf D 53} (1996) 2658, hep-ph/9507378.
\bibitem{noi} I. Dasgupta, B. A. Dobrescu and L. Randall,
    Nucl.\ Phys.\ {\bf B 483} (1997) 95, hep-ph/9607487.
\bibitem{berk}
N. Arkani-Hamed, C. D. Carone, L. Hall and 
      H. Murayama, Phys. Rev. {\bf D 54} (1997) 7032, hep-ph/9607298.
\bibitem{mass} L.~Randall, report MIT-CTP-2591 (1996), hep-ph/9612426.
\bibitem{nonren} E. Poppitz and S. P. Trivedi, report
   Fermilab-Pub-96/338-T (1996), hep-ph/9609529; \
   N. Arkani-Hamed, J.  March-Russell, and H.
   Murayama, report LBNL-39865 (1997), hep-ph/9701286; \
   N. Haba, N. Maru and T. Matsuoka, report DPNU-96-63
   (1996), hep-ph/9612468; report DPNU-97-14 (1997), hep-ph/9703250; \
   Y. Shadmi, report Fermilab-Pub-97/060-T (1997), hep-ph/9703312.
\bibitem{ads1} I. Affleck, M. Dine and N. Seiberg, Nucl.\ Phys.\ {\bf
    B 256} (1985) 557.
\bibitem{bml} R. N. Mohapatra and S. Nandi, report UMD-PP-97-082
  (1997), hep-ph/9702291.
\bibitem{su5} 
I. Affleck, M. Dine and N. Seiberg, Phys.\ Rev.\ Lett.\
   {\bf 52} (1984) 1677.
\bibitem{veld} T. A. ter Veldhuis,  Phys.\ Lett.\ {\bf
    B 367} (1996) 157, hep-th/9510121.
\bibitem{guha} J. F. Gunion and H. E. Haber, Nucl.\ Phys.\ {\bf
    B 272} (1986) 1 (Errata: UCD-92-31, hep-ph/9301205).
\bibitem{gmass} E. Poppitz and S. P. Trivedi, report
  Fermilab-Pub-97/054-T (1996), hep-ph/9703246.
\bibitem{snowmass} J. Amundson, et al,
 Report of the Supersymmetry Theory Subgroup, Snowmass, 1996,
 hep-ph/9609374.

\vfil\newpage
\bibitem{goldstino}
D. R. Stump, M. Wiest and C. P. Yuan, Phys. Rev. {\bf D 54} (1996)
   1936, hep-ph/9601362; \
S. Dimopoulos, M. Dine, S. Raby and S. Thomas,
Phys.\ Rev.\ Lett. {\bf 76} (1996) 3494, hep-ph/9601367; \
S. Dimopoulos, S. Thomas and J. D. Wells, Phys. Rev. {\bf D 54}
(1996) 3283, hep-ph/9604452; \
S. Ambrosanio, G. L. Kane, G. D. Kribs, S. P. Martin and S. Mrenna,
  Phys. Rev. {\bf D 54} (1996) 5395, hep-ph/9605398; \
  K. S. Babu, C. Kolda and F. Wilczek, Phys. Rev. Lett. {\bf 77}
  (1996) 3070, hep-ph/9605408; \
  S. Ambrosanio, G. D. Kribs and S. P. Martin, hep-ph/9703211.
\bibitem{gravitino} S. Borgani, A. Masiero and M. Yamaguchi,
  Phys. Lett. {\bf B 386} (1996) 189, hep-ph/9605222; \\
A. de Gouvea, T. Moroi and  H. Murayama, 
report LBNL-39753 (1997), hep-ph/9701244.
\bibitem{gauginos} L. Clavelli and P. W. Coulter, Phys. Rev. {\bf D
    51} (1995) 1117; \
  G. Farrar, Phys.\ Rev. Lett.\ {\bf 76} (1996) 4111,
  hep-ph/96084387; \
  S. Raby, report OHSTPY-HEP-T-97-002, hep-ph/9702299.
\bibitem{general} S. P. Martin, Phys. Rev. {\bf D 55}
  (1997) 3177, hep-ph/9608224.
\bibitem{kin} K. R. Dienes, C. Kolda and J. March-Russell,
  report IASSNS-HEP-96/100 (1996), hep-ph/9610479.
\bibitem{mu} 
B. A. Dobrescu, Phys. Rev. {\bf D 53} (1996) 6558, hep-ph/9510424; 
G. Dvali, G. F. Giudice and A. Pomarol, Nucl. Phys. {\bf B 478} 
    (1996) 31, hep-ph/9603238; 
M. Dine, Y. Nir, Y. Shirman, Phys. Rev. {\bf D 55} (1997) 1501,
 hep-ph/9607397; 
T.~Yanagida, report UT-768 (1997), hep-ph/9701394.
\bibitem{apj}
T. P. Walker, G. Steigman, D. N. Schramm, K. A. Olive and H.-S. Kang,
Astrophys. J. {\bf 376} (1991) 51.
\bibitem{mesfield} S. Dimopoulos, G. F. Giudice, A. Pomarol, 
Phys. Lett. {\bf B 389} (1996) 37, hep-ph/9607225.
\bibitem{raxion} J. Bagger, E. Poppitz and L. Randall, Nucl.\ Phys.\
  {\bf B 426} (1994) 3, hep-ph/9405345.


\end{thebibliography}
\end{document}